\documentclass[referee]{aa}

\usepackage{graphicx}
\usepackage{txfonts}
\usepackage[english]{babel}
\usepackage{url}

\usepackage{natbib}
\bibpunct{(}{)}{;}{a}{}{,} 

\begin{document}

   \title{ Experimental evidence for water formation on interstellar
dust grains by hydrogen and oxygen atoms}

  \author{F. Dulieu \inst{1}     
    \and L. Amiaud \inst{1,3}     
    \and E. Congiu \inst{1}
    \and J-H. Fillion\inst{1,4}
    \and E. Matar\inst{1}
    \and A. Momeni\inst{1,3}
    \and V. Pirronello\inst{2}
    \and J. L. Lemaire\inst{1}
          }
\authorrunning{Dulieu et al.}
\titlerunning{Water formation on interstellar dust grains}

 \institute{
    LERMA, UMR8112 du CNRS, de l'Observatoire de Paris et de l'Universit\'e de Cergy Pontoise, 5 mail Gay Lussac, 95031 Cergy Pontoise Cedex, France
    \and Dipartimento di Metodologie Fisiche e Chimiche, Universit\`a di Catania, Viale A. Doria 6, 95125 Catania, Sicily, Italy
    \and Present address: LCAM, Orsay, France
    \and Present address: Universit\'e Pierre et Marie Curie - Paris 6, LPMAA, UMR 7092 du CNRS, France
             }

\date{Received XXX; accepted YYY}

\abstract
   {The synthesis of water is one necessary step in the origin and development of life. It is believed that pristine water is formed and grows on the surface of icy dust grains in dark interstellar clouds. Until now, there has been no experimental evidence whether this scenario is feasible or not on an astrophysically relevant template and by hydrogen and oxygen atom reactions.}
   {We present here the first experimental evidence of water synthesis by such a
process on a realistic grain surface analogue in dense clouds, i.e., amorphous water ice.}
   {Atomic beams of oxygen and deuterium are aimed at a porous water ice substrate (H$_{2}$O) held at 10~K. Products are analyzed by the temperature-programmed desorption technique}
 {We observe production of HDO and D$_{2}$O, indicating that water is formed under conditions of the dense interstellar medium from hydrogen and oxygen atoms. This experiment opens up the field of a little explored complex chemistry that could occur on dust grains, believed to be the site where key processes lead to the molecular diversity and complexity observed in the Universe.}
  {}

\keywords{astrochemistry -- ISM: molecules -- ISM: dust, extinction -- methods: laboratory}

\maketitle

\section{Introduction}

Water, the spring of life \cite{brack}, is the most abundant
molecule in biological systems, and it is almost certainly of
extraterrestrial origin. Water has been detected, in gaseous or
solid form, in numerous astrophysical environments such as planets,
comets, interstellar clouds and star forming regions where strong
maser emission can be also observed \cite{ehrenfreund,dartois}.
Amorphous water ice was directly detected in dark interstellar
clouds through infra-red absorption \cite{leger}. During the
formation of stars deep inside molecular clouds, gas and dust become
part of the infalling material feeding the central object. Part of
this gas and dust grains, covered with icy mantles (mainly composed
of water), ends up in the rotating disks surrounding young stars and
forms the basic material from which icy planetesimals and later
planets, together with comets in the external regions, are formed
\cite{vanD}. While the means of delivery of water to Earth remain a
subject of debate \cite{morbidelli}, the synthesis of water in the
Universe is a fundamental link in establishing our origins. Water
molecule formation in the gas phase is not efficient enough to
reproduce the observed abundances in dark clouds, especially in its
solid form \cite{pariseb,ceccarelli}. Therefore water ice must form
directly on the cold interstellar grains and not as a condensate
after being formed in the gas phase. A complete review of the
processes involved both in the gas and solid phase has been recently
published \cite{tielensb}. It was suggested many years ago that
interstellar dust grains act as catalysts \cite{oort,vandeH}.
Starting from simple atoms or molecules such as H, O, C, N, CO,
grains are believed to be chemical nanofactories on which more
complex molecules are synthesized leading eventually to prebiotic
species produced concurrently by surface reactions and by UV photons
and cosmic rays irradiation, as already shown long ago
\cite{hagen,pirronelloa}. The most volatile species may be released
in the gas phase upon formation \cite{garrod}, while the refractory
ones remain on the grain surface, building up a so called ``dirty
icy mantle'', and at least partially may be sputtered by the heavy
component of cosmic rays \cite{johnson}. Such mantles, having a
typical thickness of a hundred monolayers, are mainly composed of
water, the most abundant solid phase species in the Universe. Under
dark cloud conditions, except for the very first monolayer that has
to grow on bare silicate or carbonaceous grains \cite{papoular},
most water molecules should be subsequently synthesized on a surface
mainly composed of water.

Chemical models including water formation on grain surfaces were
proposed years ago by \cite{tielensa}. They suggested that H$_{2}$O
formation would be initiated by H-atoms reacting with O, O$_{2}$ and
O$_{3}$, although the O$_{3}$+H pathway was considered the most
effective and O$_{2}$ would play more a catalytic role. Recent Monte
Carlo simulations \cite{cuppen} show that while the main route to
water formation on cosmic dust grains in diffuse and translucent
clouds is the reaction H~+~OH, in dense clouds the principal source
of H$_{2}$O is the reaction between H$_{2}$ and OH. This study also
emphasizes the non-negligible contribution from the
H~+~H$_{2}$O$_{2}$ reaction (H$_{2}$O$_{2}$ being a product of the
H~+~O$_{2}$ pathway) and the unusual high abundance of reactants
such as OH and O$_{3}$. Interestingly, another code by
\citet{parisea} proposed a water formation scheme where O$_{3}$
molecules react with H- or D-atoms to form OH or OD, and
subsequently the reaction H$_{2}$~+~OH/OD leads to H$_{2}$O/HDO. It
should be noted that this scheme was in part also constrained by the
observed abundances of deuterated species.

In previous laboratory works, \cite{hiraoka} succeeded in producing
water molecules from the reaction of H-~and O-atoms initially
trapped in a N$_{2}$O matrix. Very recently \cite{miyauchi}
investigated the reaction between cold H-atoms and an O$_{2}$ ice at
10~K and demonstrated the production of H$_{2}$O$_{2}$ and H$_{2}$O
molecules and estimated the efficiency of the reactions.
\cite{ioppolo} did a similar experiment but with varying O$_{2}$
substrate temperatures. They confirmed the production of
H$_{2}$O$_{2}$ and H$_{2}$O, made an estimate of the reactions
efficiency and also drew conclusions upon the temperature dependence
of the amount of species produced. These two experiments dealt with
the H$_{2}$O production pathway in which O$_2$ is the species
consumed to produce water as shown in preliminary experiments by our
group \cite{momeni,matar}

In the present study, the formation of water is studied for the
first time using hydrogen and oxygen atoms interacting on the
surface of an amorphous solid water (ASW) ice film, hence under conditions
that are much more relevant to the interstellar medium. The aim of
this first attempt to synthesize water under conditions close to
those encountered in dense clouds was to investigate how water
formation continues on the icy surfaces of cosmic grains
and to give an estimate of the efficiency of the chemical path(s)
actually active.

\section{Experimental procedures}
Experiments were performed with the FORMOLISM set-up
\cite{amiaud}. In brief, a copper sample surface whose temperature
can be controlled by computer in the 8-800~K range is maintained
under UHV conditions; on it an amorphous solid water  ice
substrate, on which water formation is studied, is prepared in two
steps. First, a 100-layer film of non-porous ASW ice is first
deposited at 120~K \cite{kimmel}, then an overlayer of 10~layers
of porous ASW ice is grown at 10~K. The underlying non-porous water
film isolates the ice layer from the copper substrate
\cite{engquist}. This double ASW ice film is annealed to 90~K
prior to each experimental run in order to avoid any further
collapse of the pores between 10~and 80~K in the subsequent
temperature-programmed desorption (TPD) experiments (see below). In
fact, the species D$_{2}$, O, O$_{2}$ used in this
experiment will be thoroughly evaporated by 90~K and thus before any
rearrangement in the porous structure of the ice template.

The annealed water ice substrate is still porous \cite{kimmel} and
mimics an amorphous ice processed by UV and cosmic rays \cite
{palumbo} that is thought to constitute the icy mantle of
interstellar grains. By using an architecture with two separate
channels \cite{pirronellob}, each consisting of a triply
differentially pumped beam line, two atomic beams of O and D are
aimed at the ASW ice substrate held at 10~K. Atoms are produced by
dissociation of O${_{2}}$ and D${_{2}}$ in microwave discharges. The
dissociation efficiency of the oxygen beam is typically 40\%,
meaning that for 100 O${_{2}}$ molecules that initially feed the
discharge, 60 O${_{2}}$ molecules and 80 O atoms will finally reach
the cold target. Trace quantities of residual gases (i.e., CO,
N$_2$, CO$_{2}$ and H$_{2}$O) are present in the beam, although
neither O$_3$ nor deuterated compounds
were detected. The D beam has a 60\% dissociation efficiency, and no
UV photons from the D$_{2}$-discharge plasma can reach the water ice
sample. The purity of the beams was checked with a rotatable
quadrupole mass spectrometer (QMS) that intercepted the incoming
beam. The O-~and D-beams have a flux of $10^{12}$ atoms/cm$^2$/s and
$5\times 10^{12}$ atoms/cm$^2$/s, respectively. After concurrent or
sequential injection of the beams, a TPD is performed at a heating
rate of 20~K/minute from 10~K to 200~K and simultaneously the
desorbing species are monitored with the QMS. The signals of mass 19
(HDO) and mass 20 (D$_2$O) presented here are corrected by
subtracting the contribution of the H$_2$O water substrate, which
naturally includes some isotopes.

\section{Results and Discussion}
\begin{figure}
   \centering
   \includegraphics[width=9cm]{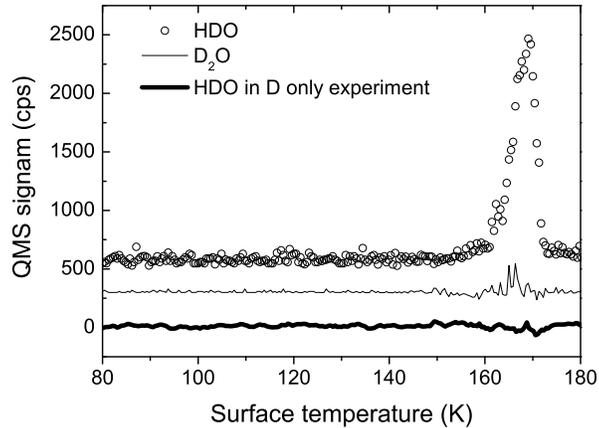}
\caption{TPD profiles of D$_2^{16}$O (thin line) and HD$^{16}$O
(circles) after irradiation of the water substrate (H$_2^{16}$O)
with D atoms, $^{16}$O atoms and $^{16}$O${_{2}}$ molecules. Thick
line: TPD profile of HD$^{16}$O, after 30 minutes of D irradiation
of the water ice substrate. The water ice substrate is held at 10~K
during all exposures. Traces are vertically shifted for clarity.
 }\label{Fig1}
   \end{figure}

Several experiments have been performed. In the first one D-atoms
were sent onto the water surface to confirm that hydrogen atoms do
not react with water molecules in the substrate to produce
deuterated water molecules (see the solid thick line in
Fig.~\ref{Fig1}), as already mentioned by \cite{nagaoka}. In
addition, we determined that D$_2$ molecules do not react with
O-atoms nor with O$_2$ molecules residing on the ASW ice surface.
This, that may seem of secondary importance, is on the contrary a
decisive result that proves that the water formation process
requires hydrogen in atomic form.

Fig.~\ref{Fig1} shows the mass spectrum recorded during the TPD
performed after the irradiation of the water substrate with D-atoms
and $^{16}$O-atoms (and $^{16}$O${_{2}}$ molecules). Water formation
clearly occurs as is testified by the presence of D$_2^{16}$O and
HD$^{16}$O peaks. These findings are not surprising although one
could expect that D$_2$O should be mainly formed, because D is used
as precursor. But it is also known that during the heating, isotope
exchange between water molecules will occur \cite{smith97}. For a
very thin layer of deuterated water, a complete
isotope exchange with the underlying H$_2$O substrate molecules
is expected around 150~K. In such a context, even if D$_2$O is formed at 10~K,
most of the deuterated water will desorb as HD$^{16}$O, as is
observed in our experiments.

Due to the mass spectrometric detection method of water molecules in
the gas phase one could wonder whether molecules synthesized by the
reaction of D-atoms and O-atoms and O$_2$ were formed during the
warm up of the whole ice substrate, a fact that would render the
result not relevant to interstellar space. However, this is
certainly not the case because reactions involving D-atoms must
proceed at temperature below 20~K. At temperatures higher than that
value the residence time of H-/D-atoms becomes exceedingly small for
reactions to occur.

In order to confirm unequivocally that water was formed starting
from the reactants deposited from the gas phase and provide
quantitative results, we repeated the experiment presented above
with the oxygen beam line fed with $^{18}$O$_2$ molecules. To be
sure that the concentration on the substrate of D-atoms on the
surface was always much higher than O-atoms and O$_2$ molecules, we
opened the oxygen beam by time-slices of 20~s every 2~minutes while
the D-beam was constantly running and irradiating the target held at
10~K. Thus, the relative concentration on the surface consists of
$\ge 10$ D-atoms per O-nucleus (in atomic or molecular form).
\cite{matar} have shown that D reacts very efficiently with O$_2$,
and that O$_2$ disappears from the surface in a one~D for one~O$_2$
fashion. In such a scenario, due to the very low mobility of O-atoms
\cite{tielensa} and O$_2$ molecules at the irradiation temperature
and in the temperature range in which H- and D-atoms remain adsorbed
on the substrate, it is reasonable to assume that O and O$_2$
surface concentration reflects the proportions in which they were
produced in the beam. O-atoms have therefore a very low probability to form O$_2$ and to encounter O$_2$
molecules to form O$_3$, especially because they react with D with a
higher probability.

\begin{figure}
\centering
\includegraphics[width=9cm]{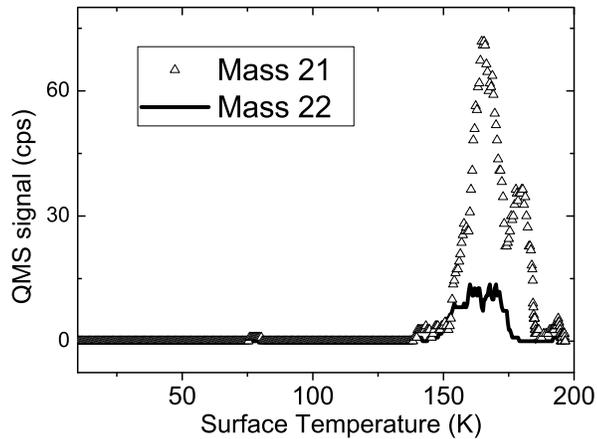}
\caption{ HD$^{18}$O (open triangles) and D${_{2}}$$^{18}$O (solid
line) desorption profiles after ten minutes of simultaneous
irradiation of the amorphous H$_2$O ice substrate with $^{18}$O-~and
D-atoms.
              }
         \label{fig2}
   \end{figure}

Fig.~\ref{fig2} shows a TPD mass spectrum obtained after a 10-minute
dose (30 time-windows of 20~s every 2~minutes) of $^{18}$O-atoms and
simultaneous D-atom irradiation of the ASW ice substrate
at 10~K. The two peaks at mass 21 and 22 can only be assigned to
HD$^{18}$O and D$_2^{18}$O and this clearly shows again that
formation of water molecules has occurred on the amorphous H$_2$O
ice substrate, namely, a realistic analogue of cosmic dust surfaces
in dense clouds.

In Table~\ref{Table:1} are shown the relative abundances (in terms
of uncorrected ion counts) of the species detected during the TPD,
and, in addition, the total number of oxygen nuclei ($^{18}$O atoms
and twice the number of $^{18}$O$_2$ molecules) that have been
exposed to the substrate during the irradiation phase. We have not
performed here any correction (except the correction from the H$_2$O
substrate).
\begin{table*}
\caption{Surface area under TPD curves of different masses after the
$^{18}$O + D experiment. Results of a statistical model are given as
a comparison.} \label{Table:1} \centering
\begin{tabular}{|l|c|c|c|c|c|c|}
  \hline
  & & & & & & \\
   & Total $^{18}$O & Mass 19 & Mass 20 & Mass 21 & Mass 22 & Mass 23 \\
  \hline

  Area (Cts.K) & 15000$\pm$1000 & 30000$\pm$2000 & 10000$\pm$500 & 1250$\pm$150 & 250$\pm$70 & 20 \\
  \hline

  Model         & 15000 & 31802 & 8076 & 1434 & 188 & 0 \\
  \hline
\end{tabular}

\end{table*}
As a preliminary remark, the signal at mass~23 is expected to be 0,
and it serves as a good indicator of the level of noise. To evaluate
the conversion efficiency of O-atoms into water molecules in our
experiment, we may evaluate the ratio $\alpha$ between the sum of
mass~20 ($^{18}$OD and H$_2^{18}$O ), mass~21 (HD$^{18}$O), mass~22
(D$_2^{18}$O) and the total amount of $^{18}$O nuclei:

$$\alpha =\frac{\mathrm{M}20 + \mathrm{M}21+ \mathrm{M}22}{\mathrm{total} \, ^{18}\mathrm{O}} = 0.76 $$

This ratio represents an upper limit to the formation of water
molecules from $^{18}$O assuming that the contribution of
D$_2^{16}$O to the signal at mass~20 is negligible, as we have
actually seen above (Fig.~\ref{Fig1}). $\alpha$, however, is not 1.
The missing $^{18}$O nuclei are likely to have partially desorbed
during the reaction or ended up in the form $^{18}$OH
radicals or even as H$_2^{18}$O$_2$ (and its deuterated
isotopologues).

The formation of water molecules through the D~+~O$_2$ pathway has
already been studied and confirmed experimentally by \cite{miyauchi,ioppolo,matar}. In the present experiment,
we can deduce a lower limit for the formation efficiency for the D+O
pathway. Actually
 60\% of the
available $^{18}$O on the surface is contained in molecules. We can
then see that, even if we assume that the molecular pathway has a
100\% efficiency, it is less than $\alpha$ and a lower limit for
water formation via the D~+~O pathway is about 0.16 ($0.76-0.60$).
Therefore, if we define the formation efficiency via atoms as the
ratio between the fraction of water molecules that certainly formed
via the D~+~O route (0.16) and the fraction of O-atoms available
(0.4), we obtain 0.4 as a lower limit for the formation efficiency.
On the other hand we can also deduce an upper limit. If we assume that O~and O$_2$ react with equal
probability with D, we can estimate that 31\% ($0.76 \times 0.4$) of
the water we have formed is the product of the D~+~O pathway and the
efficiency of water formation via atoms is equal to $\alpha$.
Considering both boundary conditions, we conclude that a reasonable
estimate for the efficiency of water formation via atoms is about
$0.5$.

In order to know if we can go deeper in the understanding of the
different pathways of water formation, we have made a very simple
statistical model which is able to reproduce correctly the data (see
Table~\ref{Table:1}). The model considers only statistical
equilibrium. For sake of clarity we give here a simple example.
 Suppose there is an initial population of 90 H$_2$O and 10 OD ,
 the final equilibrium obtained by randomly mixing the isotopes is  OD = 0.5, OH = 9.5 ,  H$_2$O = 81.225,  HDO = 8.55, D$_2$0 = 0.225.
 We have also included oxygen scrambling and calculated the statistical weight of all the final compounds. Finally,
we sum the final products by masses taking into account QMS fractionation and are able to compare with the experimental results.

The question is the following: is it possible
that the OH~+~H reaction has a activation barrier as previously
suggested in gas phase. In this case, $^{18}$OD could be mixed
with H$_2$O at higher temperature during the isotope exchange? We
first assume that the OD~+~D reaction is actually negligible
compared to the O$_2$~+~D reaction, which leads to
formation of OD and D$_2$O in equal amounts. Therefore, at the end
of the D irradiation the ratio $a=$  $^{18}$OD /D$_2^{18}$O is equal to
(40+30 =70)/30 because $^{18}$O atoms will form $^{18}$OD and $^{18}$O$_2$
molecules will form an equal amount of $^{18}$OD and D$_2$$^{18}$O (via $^{18}$O$_2$D$_2$) .
We assume only one adjustable parameter, the ratio $b$=$^{16}$O/$^{18}$O
of atoms that will exchange during the TPD and statistically equilibrate their
populations via isotope exchanges.

 We find a best fit (Table~\ref{Table:1}) for the only free parmeter $b=15$.
This means that the overlayer of products ($^{18}$OD and D$_2^{18}$O) is mixed
with 15~layers of the H$_2$O substrate and this is reasonably consistent with
previous results. \cite{smith00} found an efficient mixing of about
50~ML of both isotopes, although they used a lower heating ramp and
a different amorphous ice. If one believes in
our simple statistical model, it is also possible to explain our data starting with an high population of
OD, and therefore that an activation
barrier for the OD~+~D reaction exists at 10~K.
Indeed, the isotope exchange blurs the chemical pathways, and the
data provided here give not enough constraints to make a firm
conclusion about which pathway is more efficient. None the less, on
average, we have estimated that water formation via the D~+~O
pathway is highly efficient (about 50\%) and certainly is
responsible for producing a significant fraction of water molecules.
More laboratory investigations should be performed and we actually
plan to do that by combining TPD and infra-red spectroscopy.

The relevance of our results to astrochemistry is clear and gives
strong experimental support to simulations of the formation and time
scale of growth of water ice mantles in dense clouds. For the first
time it is demonstrated on experimental grounds that water molecules
can form on an amorphous water ice substrate under interstellar
conditions (i.e., through surface reactions between atomic hydrogen
and atomic and molecular oxygen on ASW ice), allowing the growth of
icy mantles that are observed in dense clouds. Also, in our
experiments, as occurs on interstellar grains, the formation of
water ice via reactions between hydrogen and oxygen atoms suffers
from the competition between the formation of molecular hydrogen and
molecular oxygen and probably this competition might reflect in the
depth profile inside the ice layer and in the time evolution of the
mantle itself.
Several issues still need to be investigated: the efficiency at
various coverages and with an abundance ratio between hydrogen and
oxygen atoms closer to the interstellar one; the interaction of the
atoms that land from the gas phase and those belonging to the
surface; other chemical pathways that contribute to the formation of
water and so on.  Besides all these investigations,
of course, still to prove is how the first monolayer of water
molecules may form efficiently on an amorphous silicate and/or and
amorphous carbon layer. If it can not form efficiently enough, it
will be compulsory to accept that the very first monolayer has to be
built by water molecules formed by gas-phase reactions and then
accreted on the bare refractory grain surface; a fact of major
interest inside dense clouds near the threshold of observability of
icy mantles.  These and other issues will be addressed in
forthcoming papers.

\section{Conclusions}
We presented the first laboratory attempt to reproduce the formation
of water molecules on of realistic space analogue of grain surfaces
in dense molecular clouds. By exposing O-~and D-atoms to an
amorphous water ice substrate held at 10~K deuterated water
molecules were formed with a high efficiency ($\sim$ 50~\%).

\begin{acknowledgements}
We acknowledge the support of the French National PCMI program funded
by the CNRS, as well as the strong financial support from the Conseil
R\'egional d'Ile de France through a SESAME program (No. E1315) and
from the Conseil G\'en\'eral du Val d'Oise. We also thank E. herbst, S. Leach, H. J. Fraser and P. Cernicharo for their
comments on an early version of this work.
\end{acknowledgements}

\bibliographystyle{aa} 

\end{document}